\documentclass[preprint,12pt]{revtex4-1}
\usepackage{graphicx}
\usepackage{bm}
\begin{document}

\title{Characterization of Interface Traps in SiO$_2$/SiC Structures Close to the Conduction Band by Deep-Level Transient Spectroscopy}

\author{Tetsuo Hatakeyama}
\email{tetsuo-hatakeyama@aist.go.jp}
\author{Mitsuru Sometani}
\author{Yoshiyuki Yonezawa}
\author{Kenji Fukuda}
\author{Hajime Okumura}
\affiliation{National Institute of Advanced Industrial Science and Technology, Centeral 2-13,
 1-1-1 Umezono, Tsukuba, Ibaraki 305-8568, Japan}
\author{Tsunenobu Kimoto}
\affiliation{
Department of Electronic Science and Engineering, Kyoto University, Katsura, Nishikyo, Kyoto, 615-8510, Japan
}%

\begin{abstract}
The effects of the oxidation atmosphere and crystal faces on the interface-trap density
was examined by using constant-capacitance deep-level transient spectroscopy
to clarify the origin of them. 
By comparing the DLTS spectra of the low-mobility interfaces oxidized in a N$_2$O atmosphere
with those of the high-mobility interfaces on C-face oxidized in a wet atmosphere, 
it was found that a high density of traps are commonly observed around the energy of 0.16 eV 
from the edge of the conduction band ($C1$ traps) in low-mobility interfaces irrespective of
crystal faces.  It was also found that the generation and elimination of traps specific 
to crystal faces: (1) the $C1$ traps 
can be eliminated by wet oxidation only on the C-face, and (2) the $O2$ traps (0.37 eV) can be observed
in the SiC/SiO$_2$ interface only on the Si-face.
The generation of $O2$ traps on the Si-face and the elimination of $C1$ traps
on the C-face by wet oxidation may be caused by the oxidation reaction specific to the crystal faces.
\end{abstract}
\maketitle
\newpage
\section{Introduction}
SiC metal--oxide--semiconductor field-effect transistors (MOSFETs) are regarded as promising candidates 
for the next-generation high-voltage electrical power switches owing to the high critical electric field 
of SiC \cite{palmour, cooper, baliga}.
However, the low mobility in the SiC/SiO$_2$ interfaces hinders the potential performance of SiC MOSFETs.
Thus, the improvement in the mobility in the SiC/SiO$_2$ interfaces is a central issue in the research and
development of SiC MOSFETs.
It was presumed that the traps  in the SiC/SiO$_2$ interfaces  are closely related to the degradation 
in mobility \cite{afanasiev_pss}.
In 2000, Saks and Agarwal clearly showed that the low mobility in the SiC/SiO$_2$ interfaces 
is caused by the trapping of electrons at the high-density interface traps on the bases of the Hall effect 
measurements of SiC MOSFETs\cite{saks2000}.
They showed that most of the inversion electrons induced by the gate voltage were trapped by interface traps
by comparing the free carrier density in an interface obtained by Hall measurements with the
 total number of inversion electrons. 
They also pointed out that the Coulombic scattering by the trapped electrons may dominate
the inversion electron transport by examining the temperature dependence of the Hall mobility.
Later, detailed studies on the inversion electron transport of various types of SiC MOSFETs 
using Hall measurements confirmed the above-described degradation mechanism in mobility \cite{tilak_pss,dhar_jap}.\par
Therefore, great efforts have been focused on reducing the interface trap density
to improve mobility by examining the gate-oxidation and post-gate-oxidation annealing processes in detail.
In recent years, annealing or oxidation in a nitric oxide (NO) or nitrous oxide (N$_2$O) atmosphere, 
which is hereinafter collectively referred to as oxynitridation, 
has been used to reduce the high density of interface traps \cite{Jamet, Chung, Rozen}.  
The optimized oxynitridation process reduces the interface trap density  ($D_{\mathrm it}$)
evaluated by using the conventional Hi-Lo method \cite{nicollian_hilo}
down to less than 10$^{12}$ cm$^{-2}$/eV at $E_{\mathrm C}-E$ = 0.2 eV, where $E_{\mathrm C}$
 and $E$ are referred to as the conduction-band edge and energy, respectively \cite{Suzukino}.  
However, the effect of oxynitridation on the mobility is limited. 
In fact, the mobility in the SiC/SiO$_2$ interfaces fabricated by using oxynitridation is typically
approximately 30 cm$^2$/(Vs)  \cite{Suzukino,Rozenno}. 
Another way to improve the channel mobility is to combine the use of the C-terminated face (C-face) instead 
of the Si-terminated face (Si-face) along with annealing or oxidation in a wet atmosphere.
 The typical mobility in the SiC/SiO$_2$ interfaces fabricated on the C-face
 by using wet oxidation is approximately 90 cm$^2$/(Vs) \cite{fukuda, Suzukiwet}. 
However, the cause of the relatively low mobility of the oxynitrided interface
 could not yet be identified.
The densities of interface traps characterized by using the conventional 
Hi-Lo method are not correlated with the mobilities between these two types
 of samples \cite{Suzukino, Suzukiwet}; thus, it seems that $D_{\mathrm it}$
 would not be the cause of the relatively low mobility of the oxynitrided interface.\par
 
 The authors reported that the  $D_{\mathrm it}$ close to the conduction band in SiC/SiO$_2$ interfaces
 fabricated using oxynitridation was much higher than that of SiC/SiO$_2$ interfaces fabricated using wet oxidation on C-face by using constant-capacitance deep-level transient spectroscopy (CCDLTS) characterization. \cite{HatakeyamaDLTS}. They concluded that 
the low mobility in SiC/SiO$_2$ interfaces oxidized in a N$_2$O atmosphere on C-face 
should be caused by trapping electrons in the high density of the traps
close to the conduction band.
In this study, the effects crystal faces and oxidation atmosphere on the traps in SiC/SiO$_2$ interfaces
was examined to confirm that the high density of the traps close to the conduction band are
the common cause of the low mobility in SiC/SiO$_2$ interfaces irrespective of the crystal faces. 
Further, to elucidate the origin of the traps in SiC/SiO$_2$ interfaces,
the properties of the identified traps close to the conduction band are
discussed according to the dependence of the CCDLTS spectra on
the crystal face and oxidation condition.\par
\section{Experimental Methods}
The samples characterized in this study were MOS capacitors on the
C-face $(000\overline{1})$ or Si-face $(0001)$ of 4H-SiC n-type epitaxial wafers.
The density of nitrogen in the epitaxial layer was approximately $1\times 10^{16}$ cm$^{-3}$. 
The SiC/SiO$_2$ interfaces of the MOS capacitors were fabricated by using 
the following gate-oxidation processes: 
(1) oxidation in an O$_2$ atmosphere at 1250 ${}^\circ\mathrm{C}$,
 followed by wet oxidation at 900 ${}^\circ\mathrm{C}$,
 followed  by H$_2$ anneal at 800${}^\circ\mathrm{C}$ on the C-face (DWHC); 
 (2) oxidation in a N$_2$O atmosphere at 1250 ${}^\circ\mathrm{C}$,
 followed by a H$_2$ anneal at 1000 ${}^\circ\mathrm{C}$ on the C-face (NHC);
 (3) oxidation in an O$_2$ atmosphere at 1250 ${}^\circ\mathrm{C}$,
 followed by wet oxidation at 900 ${}^\circ\mathrm{C}$ on the Si-face (DWS); and 
(4) oxidation in an O$_2$ atmosphere at 1250 ${}^\circ\mathrm{C}$,
 followed by post-oxidation annealing in a N$_2$O atmosphere at 1250 ${}^\circ\mathrm{C}$,
 followed by  H$_2$ anneal at 800 ${}^\circ\mathrm{C}$ on the Si-face (DNHS).
The thickness of the oxide layer is approximately 50 nm, and
the gate electrode is aluminum.
The mobilities of the MOSFETs fabricated  by using the processes of DWHC, NHC, DWS, and DNHS
 are approximately 80 cm$^2$/(Vs), 30 cm$^2$/(Vs),  8 cm$^2$/(Vs), and 30 cm$^2$/(Vs), respectively 
\cite{Suzukiwet, Suzukino,hatakeyama_mobility}. \par
CCDLTS spectra were obtained by measuring the transient voltage signal generated by a feedback loop 
to maintain the capacitance at a constant value during the measurement of MOS capacitors
in the temperature range from 80 K to 400 K.
The pulse and reverse bias voltage were approximately 6 V and $-1$ V, respectively. 
The capacitance at the reverse bias was kept constant during the temperature scan.
For the analysis of the transient voltage signal at each temperature,
a deep-level transient Fourier spectroscopy (DLTFS) technique was used \cite{Weiss, Weissphd}.
\begin{figure}[tbp]
\begin{center}
\includegraphics[width=7cm]{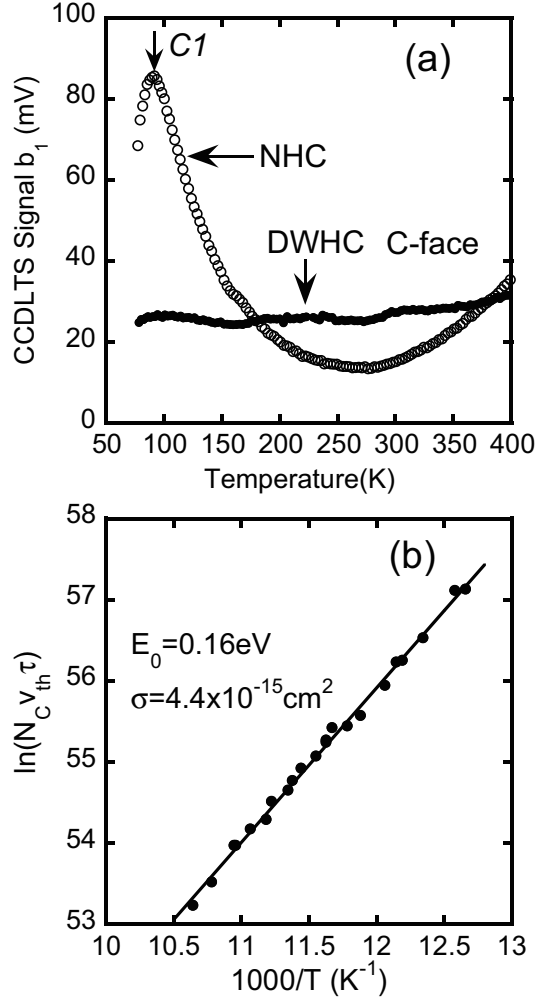}
\end{center}
\caption{%
(a) Comparison of CCDLTS spectra between a DWHC sample and an NHC sample. 
The horizontal axis is the first order of the sine coefficient of the DLTFS signal (b1) \cite{Weiss, Weissphd}. 
(b) An Arrhenius plot for the peak at approximately 100 K ($C1$) in the CCDLTS spectrum for the NHC sample.
}%
\label{fig_1}
\end{figure}
\section{Results and Discussion}
First, we examined the CCDLTS spectra of MOS capacitors on C-face to clarify the cause of 
the low mobility of the oxynitrided interface \cite{HatakeyamaDLTS}.
Figure \ref{fig_1} (a) shows the comparison of the CCDLTS 
spectra between a DWHC sample, the interface of which exhibits a high mobility, 
and an NHC sample, the interface of which exhibits relatively low mobility.
In Fig. \ref{fig_1} (a), the horizontal axis is the first order of the sine coefficient 
of the DLTFS-signal (b1) with a period width of 205 ms and a recovery time of 
4 ms \cite{Weiss, Weissphd}.
A peak was observed at approximately 100 K in the CCDLTS spectrum for the NHC sample.
We refer to this peak as $C1$ for which an Arrhenius-plot analysis was carried out, and 
the results was presented in Fig. \ref{fig_1} (b).
The obtained energy of the traps that comprise the $C1$ peak ($C1$ traps)
was estimated to be 0.16eV. 
The capture cross-section of the $C1$ traps was estimated to be 4 $\times$10$^{−15}$cm$^{−2}$.
In contrast, the CCDLTS spectrum for the DWHC sample is almost constant.
Especially, the CCDLTS signal at approximately 100 K for the the DWHC sample is
one-fourth of that for for the NHC sample.
This means that the areal density of $C1$ traps of the DWHC sample is approximately 
one-fourth of that of the NHC sample.
We conclude that the low mobility in the SiC/SiO$_2$ interfaces fabricated by the NHC process is
caused by the high density of $C1$ traps for the following reasons: 
(1) the interface mobility is inversely correlated with the density of  $C1$ traps,
and (2) the interface mobility degradation mechanism
proposed by Saks\cite{saks2000} can be applied to the high density of $C1$ traps
because the energy level of the $C1$ traps (0.16 eV) is located above the Fermi energy 
at the onset of the formation of the inversion layer 
(approximately 0.2 eV from the edge of the conduction band at room temperature). 
Consequently,  the $C1$ traps are not filled by electrons at
the onset of the formation of the inversion layer; thus some of the inversion electrons
are captured when the gate voltage exceeds the threshold voltage, 
which leads to a degradation in the interface mobility, as described in the introduction.\par
\begin{figure}[tbp]
\begin{center}
\includegraphics[width=7cm]{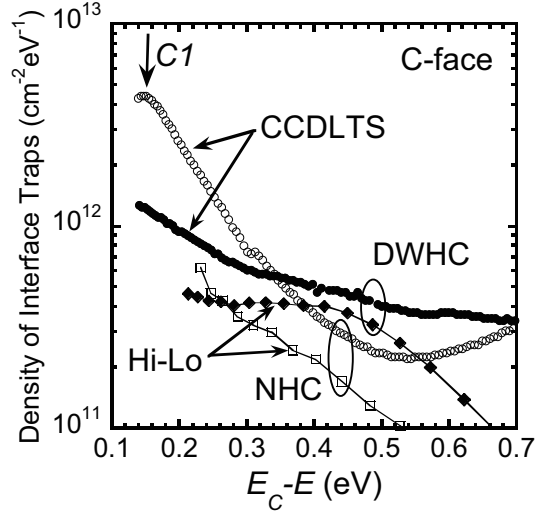}
\end{center}
\caption{%
$D_{\mathrm{ it}}(E)$ for the DWHC and NHC samples transformed from the CCDLTS spectra. 
$D_{\mathrm{ it}}(E)$ characterized via the Hi-Lo method  (100 kHz for high frequency) 
are also shown for comparison. 
}%
\label{fig_2}
\end{figure}
The CCDLTS spectra can be transformed into the energy  distribution of  the density of the 
interface traps ($D_{\mathrm it}(E)$) with the following two assumptions:
(1) $D_{\mathrm it}(E)$ depends only weakly on the energy, and
(2) the capture cross section does not depend on the energy and temperature.
Figure \ref{fig_2} shows the energy distributions for the DWHC and 
NHC samples calculated from the CCDLTS spectra.
In the calculation of $D_{\mathrm it}(E)$, the capture cross sections
for the DWHC and NHC samples are assumed to be 4$\times$10$^{−15}$cm$^{−2}$
and 1$\times$ 10$^{−15}$cm$^{−2}$, respectively.
For comparison, $D_{\mathrm{ it}}(E)$ calculated via the Hi-Lo method are also shown.
$D_{\mathrm{ it}}(E)$ for the NHC sample steeply increases as the energy become close to 
the edge of conduction band, whereas that for the DWHC sample gradually increases.
As a result, $D_{\mathrm{ it}}(E)$ close to the conduction band for the NHC sample 
is larger than that for the DWHC sample. 
This large $D_{\mathrm{ it}}(E)$ close to the conduction band for the NHC sample
corresponds to the $C1$ traps, and they degrade the MOS mobility for 
the reason as described above.
In contrast, the small $D_{\mathrm{ it}}(E)$ close to the conduction band for the DWHC sample
results in a relatively large interface mobility of 80 cm$^2$/(Vs). \par
\begin{figure}[tbp]
\begin{center}
\includegraphics[width=8cm]{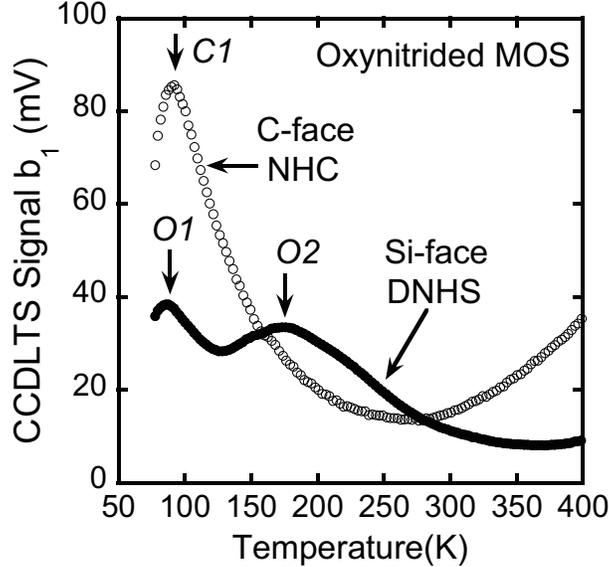}
\end{center}
\caption{%
Comparison of CCDLTS spectra between  the NHC sample and the DNHS sample.}
\label{fig_3}
\end{figure}
In Fig. \ref{fig_2}, we also present the difference of the energy 
distributions of $D_{\mathrm{ it}}(E)$ characterized from the CCDLTS spectra and 
those characterized according to the Hi-Lo method \cite{nicollian_hilo},
where high-frequency C-V characteristics are measured at 100 kHz.
It can be seen that the Hi-Lo method underestimates  $D_{\mathrm{ it}}(E)$ compared to
those estimated from CCDLTS spectra.
This is because the frequency of the high-frequency capacitance measurement (100 kHz ) 
is not high enough to measure the real “high-frequency capacitance” \cite{yoshioka_psi}. 
We note that the interface traps are modeled as a series connection of the resistance 
and the capacitance in the equivalent circuit of a MOS capacitor \cite{nicollian_hilo}. 
Accordingly, the interface traps have cut-off frequencies. 
To measure the real “high-frequency capacitance”, the frequency of the C-V measurement
should be higher than the cut-off frequency of the traps, 
which exponentially increases as the energy of traps becomes close to the edge 
of the conduction band \cite{nicollian_hilo}. Therefore, the measured “high-frequency capacitance”
is overestimated at the energy close to the edge of the conduction band,
which leads to the underestimation of $D_{\mathrm{ it}}(E)$. 
As for the deep traps ($>$ 0.5 eV), the capacitance measurements tend to 
be carried out in a non-equilibrium state, which 
also leads to the underestimation of $D_{\mathrm{ it}}(E)$.
As for the NHC sample, the pile up of the nitrogen atoms at the SiC/SiO$_2$ interface\cite{haney_2013}
may cause a deviation in the estimate of trap energy in the C-V measurements. 
In summary, the $D_{\mathrm{ it}}(E)$ characterization of
the SiC/SiO$_2$ interfaces via the Hi-Lo method at room temperature has
 a numbers of problems; thus, it should be avoided. \par
Hereafter, we discuss the difference between the CCDLTS spectrum for
MOS capacitors on the C-face and that for MOS capacitors on the Si-face
to consider the origin of the $C1$ traps and  other defects at the SiC/SiO$_2$ interfaces.
Figure \ref{fig_3} shows a comparison of the CCDLTS spectrum between the oxynitried MOS
capacitor on the C-face (the NHC sample) and the one on the Si-face (the DNHS sample). 
We found two peaks ($01$ and $O2$) in the CCDLTS spectrum for the DNHS sample,
as shown in Fig.\ \ref{fig_3}.
From an Arrhenius-plot analysis, the energies of the $O1$ traps and $O2$ traps 
are estimated to be 0.14 eV and  0.37 eV, respectively.
These peaks were also reported by Basile and his coworkers \cite{Basileno}.
It should be noted that the energy of the $C1$ trap at the SiC/SiO$_2$ interface on the C-face 
is almost equal to that of the $O1$ trap on the Si-face.
On the other hand,  the $O2$ peak in the CCDLTS spectrum is 
specific to the SiC/SiO$_2$ interface on the Si-face.
The absence of  the $O2$ peak in the CCDLTS spectrum of the SiC/SiO$_2$ interface on the C-face
means that the density of  $O2$ traps on C-face is, at least, negligible compared with that of the $O1$
traps.  This information on the dependence of the trap densities on the crystal faces provides an insight 
into the origin and formation mechanism of traps in the SiC/SiO$_2$ interface.\par
Here, we review the structure of the SiC/SiO$_2$ interface on the Si-face and C-face.
For the SiC/SiO$_2$ interface on the Si-face, uppermost Si atoms, which terminate the SiC layer,
are connected to the O atoms in the SiO$_2$ layer \cite{deak_iop2007, 
Ohnuma_2007Si, devynck2011, devynckphd, xshen_jap2013}.
For the SiC/SiO$_2$ interface on the C-face, it may be reasonable to assume that
the uppermost C atoms are connected to the O atoms in the SiO$_2$ layer.
However, first-principles molecular-dynamics calculations showed that this interface structure 
is not stable \cite{Ohnuma_2009C}. Consequently, it is believed that the Si atoms in the SiO$_2$ layer are
connected to the uppermost C atoms in SiC \cite{Ohnuma_2009C,xshen_jap2013}.
One of this type of SiC/SiO$_2$ structure was proved to be stable according to first-principles 
molecular-dynamics calculations \cite{Ohnuma_2009C}. 
Whatever else it might be, the SiC/SiO$_2$ interface on the C-face may be more unstable 
than that on the Si-face.
This may cause the high oxidation rate of the C-face, which is ten times higher than 
that of the Si-face \cite{ysong_jap2004}. 
Further, the oxidation mechanism may be different between C-face and Si-face \cite{xshen_jap2013}. 
We speculate that the generation of $O2$ traps on the Si-face may be due to
the oxidation mechanism specific to the Si-face.\par
Basile and his co-workers concluded that the $O1$ and $O2$ traps are defects in the oxide 
on the base of  the comparison of CCDLTS spectra between MOS structures on the Si-face of 4H-SiC 
and those on the Si-face of 6H-SiC \cite{Basileno}.
These traps correspond to near-interface oxide traps (NIT), which was first reported by Afanasev 
and his coworkers in 1997
on the bases of  the experiments on phton-stimulated tunneling of trapped electrons (PST) \cite{afanasiev_pss}.
Their PST measurements on MOS structures on 4H-SiC and 6H-SiC showed a barrier height of 2.8 eV, 
which corresponds to  an energy for NIT levels at approximately $E_{C}-0.1$ eV, where $E_C$ is the energy
of the edge of the conduction band of 4H-SiC.
The idea of NIT was also supported by thermally stimulated current measurements
using MOS structures on 4H-SiC and 6H-SiC \cite{Rudenko2005545}.
In consideration of these reports, the $C1$ traps on the C-face, the $O1$ traps and $O2$ traps on the Si-face
are likely to be oxide traps. Further, the $C1$ traps on the C-face are likely to be same as the $O1$ traps
on the Si-face because the energy of each of them is almost the same.
We presume that the origin of a $C1$ trap is a carbon dimer or a single carbon defect in SiO$_2$
by comparing the energy of the $C1$ traps with the charge transition energy of a point defect in
SiO$_2$ on the basis of  first-principles calculations \cite{deak_iop2007,devynck2011new}.\par
\begin{figure}[tbp]
\begin{center}
\includegraphics[width=8cm]{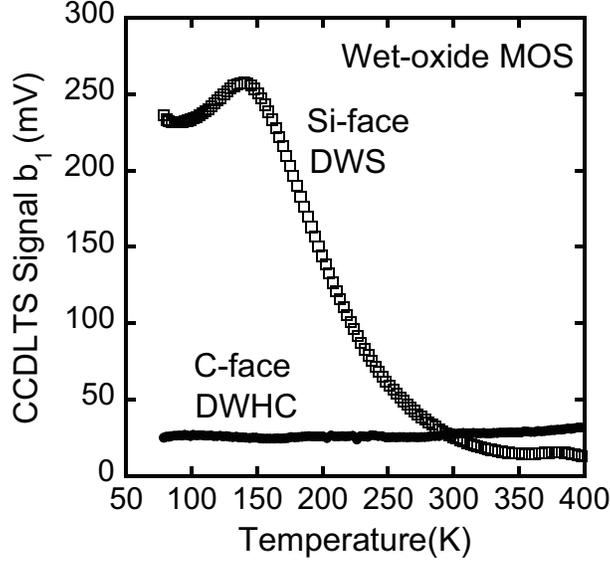}
\end{center}
\caption{%
Comparison of CCDLTS spectra of SiC/SiO$_2$ interfaces oxidized in a wet atmosphere on the C-face
(a DWHC sample) with that on Si-face (a DWS sample).
}%
\label{fig_4}
\end{figure}
A comparison of CCDLTS spectra of SiC/SiO$_2$ interfaces
oxidized in a wet atmosphere on the C-face (a DWHC sample) with that on the Si-face (a DWS sample)
is shown in Fig. \ref{fig_4}.
The CCDLTS spectrum of the DWS sample (Si-face) is much larger than that of the DWHC sample (C-face).
This difference in the characteristics of the CCDLTS spectra reflects the difference in 
interface mobilities between the C-face and the Si-face (DWHC: 80 cm$^2$/(Vs), DWS: 8 cm$^2$/(Vs)).
Figure \ref{fig_4} shows that the $C1$ traps are passivated or removed only on the C-face. 
If we assume that the $C1$ trap is an oxide trap, possible mechanisms
for the elimination of the $C1$ traps from the interface can be narrowed down.
First, we exclude the possibility of  the acceleration of the decomposition of the $C1$ traps
by wet oxidation
because the $C1$ traps could be removed by wet oxidization also on the Si-face if this mechanism works.
Therefore, it is natural to think that the difference in the density of the $C1$ traps 
between the C-face and the Si-face is due to the difference in the defect-generation rate during
wet oxidation. As described above, the  structure of  the SiC/SiO$_2$ interface on C-face
may be totally different from that on Si-face. It is certain that the oxidation mechanism
in a wet atmosphere is different between the C-face and the Si-face and that the difference in
the oxidation mechanism causes the difference in defect-generation rate at the 
oxidation front.
A more detailed investigation of wet oxidation of SiC from first principles
is needed to clarify the mechanism of removal of the $C1$ traps.
\section{Conclusions}
We used CCDLTS measurements to characterize and compare $D_{\mathrm{it}}(E)$ close to the edge of the
conduction band for SiC/SiO$_2$ interfaces on the Si-face and C-face fabricated using two techniques:
oxynitridation and wet oxidation. 
The results showed that the $D_{\mathrm{it}}(E)$ close to the edge of the
conduction band for the SiC/SiO$_2$ interface on the C-face and Si-face 
fabricated by using oxynitridation was much higher 
than that on C-face fabricated by using wet oxidation. 
The high value of  $D_{\mathrm{it}}(E)$ close to the edge of the conduction band of oxynitridated samples
is due to the $C1$ traps, 
which are likely to be the main cause of the low interface mobility.
The origin of  the $C1$ traps is likely to be the carbon-related defects in the oxide, 
which are common in the SiC/SiO$_2$ interfaces on the C-face and Si-face.  
We found $O2$ traps in the SiC/SiO$_2$ interface only on the Si-face.
We also found that $C1$ traps in the interface can be eliminated only on the C-face by wet oxidation. 
It is presumed that the generation of $O2$ traps 
in the interface on the Si-face and the elimination of $C1$ traps
in the interface on the C-face by wet oxidation are caused by the oxidation
reactions specific to the crystal faces, which are caused by the different atomic structures of the SiC/SiO$_2$ interface between the Si-face and the C-face.

\begin{acknowledgments}
We thank Dr. S. Weiss and Dr. L. Cohausz at Phys Tech GmbH and Dr. H. Okada at Kobelco research, Inc. for their help with DLTS measurements. This research was supported by a grant from the Japan Society for the Promotion of Science (JSPS) through the Funding Program for World-Leading Innovative R \&D on Science and Technology (FIRST Program), under the aegis of the Council for Science and Technology Policy (CSTP).
\end{acknowledgments}
\bibliography{DLTS_htk}

\end{document}